\documentclass[epj]{svjour}

\usepackage{latexsym}
\usepackage{makeidx}         
\usepackage{graphicx}        
\usepackage{multicol}        
\usepackage[bottom]{footmisc}
\usepackage{subfigure}
\usepackage{setspace}

\makeindex             

\begin{document}

\title*{Studying the emergence of invasiveness in tumours using game theory}
\author{David Basanta\inst{1*}\and
Haralambos Hatzikirou\inst{1}\and
Andreas Deutsch\inst{1}}
\institute{Zentrum f\"ur Informationsdienste und Hochleistungsrechnen, Technische Universit\"at Dresden, Dresden, Germany\\
\texttt{contact email:david.basanta@moffitt.org}\\\\
\texttt{Cite:D. Basanta, H. Hatzikirou and A. Deutsch. Studying the emergence of invasiveness in tumours using game theory European Physical Journal B 63 (3)	 , 393-397	 (2008) }}

\abstract{Tumour cells have to acquire a number of capabilities if a neoplasm is to become a cancer. One of these key capabilities is increased motility which is needed for invasion of other tissues and metastasis. This paper presents a qualitative mathematical model based on game theory and computer simulations using cellular automata. With this model we study the circumstances under which mutations that confer increased motility to cells can spread through a tumour made of rapidly proliferating cells. The analysis suggests therapies that could help prevent the progression towards malignancy and invasiveness of benign tumours. }

\maketitle

\section{Introduction}
Carcinogenesis is the process that describes the transformation of healthy cells into cancer cells. Cancer has long been recognised as an evolutionary disease \cite{nowell:1976}. It is also a disease in which the environment (other cells in the tissue, nutrients, pH, etc) determines which genetic mutations lead to phenotypes that spread through the population.
Hanahan and Weinberg \cite{hanahan:2000} have described six capabilities that tumour cells have to acquire during carcinogenesis for a tumour to become a cancer, that is, a tumour capable of invasion and metastasis. These capabilities include: unlimited replicative potential, environmental independence for growth, evasion of apoptosis, angiogenesis and invasion. 
This paper focuses on how a tumour made of rapidly proliferating cells acquires invasiveness with some of the cells becoming capable of motility. We call proliferative cells those cells that do not rely on the environment for growth, which in many cases involves mutations in genes such as APC, K-RAS and TGF. The motile cells are those tumour cells that, through abnormal regulation of the synthesis of cadherins and integrins, have an increase in their motility and invasiveness compared to other cell types \cite{mareel:1998,foty:2004}. This transition is a prerequisite for invasion and is key in the progression of a tumour towards malignancy.

Tumours are the result of somatic evolution: the mutations that promote tumour initiation lead to an ecosystem in which different tumour cell phenotypes compete for space and resources \cite{Merlo:2006}. Evolutionary game theory is an appropriate modelling tool in which to frame tumourigenesis \cite{gatenby:2003} and in which to study the possible equilibria between two or more phenotypes under different microenvironmental circumstances. In this paper we also describe an agent-based model using a cellular automaton. This alternative model, in which the agents can be either proliferative or motile tumour cells, has been designed to mirror, as closely as possible, the game theoretical model but whereas the first model is suitable for analysis, the simulations provided by the second have been used to study the influence of space on the results. Both models show that abundance of resources reduces the likelihood of success of motile cells. They also show that typically, there will be many circumstances that could lead to polyclonal tumours that would include both: proliferative and motile cells.

\section{Previous studies}
Game theory (GT), first formalised by von Neumann and Morgenstern \cite{Neumann:1953}, has a long tradition in the field of economy \cite{Merston-Gibbons:2000}. It has also been successfully applied to evolutionary biology \cite{Maynard:1982,Hofbauer:1998}. Its use in the field of theoretical medicine is much more recent although its future looks quite promising \cite{gatenby:2003}. GT is a powerful tool when studying the interactions between a number of entities called players in which the fitness of each player depends not only on what the player decides to do (its strategy) but also on what the other players do. In evolutionary GT the strategy of an individual is an aspect of its phenotype that affects its behaviour in the game. One important concept in evolutionary game theory is that of an evolutionary stable strategy (ESS). A strategy is ESS if, when adopted by a population, it cannot be successfully invaded by an alternative strategy as a result of evolution. 
An overview of the application of GT in cancer research can be found in \cite{Basanta:2008}. The first use of GT in the field of cancer research was described by Tomlinson and Bodmer \cite{tomlinson:1997a,tomlinson:1997b}. Their models are based on pairwise interactions between different types of tumour cell phenotypes in the context of several cancer related problems such as angiogenesis, evasion of apoptosis or the production of cytotoxic factors. In these problems, game theory is employed to analyse the circumstances that lead to polymorphism that is, coexistence of two phenotypes. Subsequent research carried out by Bach and colleagues extended this idea to interactions between three players in the angiogenesis problem \cite{bach:2001} and also studied the effect of spatial dynamics on the equilibria \cite{bach:2003}. Gatenby and Vincent adopted a game theory approach heavily influenced by population dynamics to investigate the influence of the tumour-host interface in colorectal carcinogenesis \cite{gatenby:2003a,Gatenby:2005} and suggest therapeutic strategies \cite{gatenby:2003b}. More recently Mansury and colleagues \cite{mansury:2006} have produced a modified game theory approach  in which they study how the interactions between two types of phenotypes, proliferative and motile, in a tumour affect a number of features of the tumour growth dynamics. This model does not contemplate evolutionary dynamics and thus cannot address questions related to the effect of model parameters on the coexistence of different phenotypes in a polymorphic tumour. Recently, Axelrod and colleagues have suggested, although never demonstrated, how GT could be used to study the possible coperation of different tumour sub-populations during tumour progression \cite{Axelrod:1981,Axelrod:2006}.

Cellular automata (CA) are another invention of von Neumann \cite{Neumann:1966} who, together with Stanislaw Ulam, created them to study self replication. CA are based on the concepts of discrete space and time and are well suited to model processes at the cellular level \cite{deutsch:2005}. CA have been extensively used to model many aspects relevant in cancer research  \cite{Andreas-Deutsch-Editor:2007,Moreira:2002} such as early tumour growth \cite{mansury:2006,Kansal:2000}, invasion\cite{Ribba:2004,Wurzel:2005,Hatzikirou:2005,Hatzikirou:2007b}, angiogenesis \cite{Alarcon:2003,Merks:2006}  and evolution \cite{Spencer:2006,Anderson:2006}. 

\section{The game theory model}
The model assumes a very simple microenvironment inhabited initially only by proliferative cells mimicking the situation in many benign tumours  \cite{nowell:1976}. The motile phenotype can appear as a result of one or more genetic mutations. The interactions between these two phenotypes are characterised in table \ref{tab:payoffTable}. Parameter \textit{b} is the base payoff that corresponds to a cell that does not need to share resources (glucose, oxygen, etc) with any one else. The cost of motility \textit{c} represents the fitness loss incurred by any motile cell that at a given moment chooses to move to a neighbouring location. This cost could represent for instance the fact that cells that are capable of motility cannot divide while moving \cite{giese:1996,giese:2003}.  When a motile cell finds a proliferative cell then the proliferative cell stays and uses all the resources of that location. The motile cell moves to a different location where it is assumed that it will find a similar amount of resources. When two motile cells meet then one of them stays and takes the resources whereas the other moves to another location. When two proliferative cells interact they have to share the resources of the location in which they find themselves. Given the payoffs shown in table \ref{tab:payoffTable}, the Bishop-Cannings theorem \cite{Maynard:1982} can be used to obtain the mixed strategy that would be an evolutionary stable strategy (ESS). A mixed strategy is a strategy in which one of a number of pure strategies can be used with a given probability. If the pure strategies are to be motile or to be proliferative then a mixed strategy could be 0.6 implying that the cell would play the strategy of motility with probability 0.6 and the strategy proliferative with 0.4. A mixed strategy in this game could also correspond to a population composed of more than one phenotype.

\begin{table}[h]        
	\begin{center}
	\caption{Payoff table that represents the change in fitness of a tumour cell with a given phenotype interacting with another cell. The base payoff in a given interaction is b and the cost of moving to another location with respect to the base payoff is c.}
	\begin{tabular}{|c|c|c|} \hline
	& Proliferative                &      Motile     \\ \hline
	Proliferative           &       $\frac{1}{2} $ &    b-c       \\ \hline       
	Motile             & b  & $b-\frac{c}{2}$       \\ \hline
	\end{tabular}  
	\label{tab:payoffTable}
	\end{center}
\end{table}

The genetic polymorphism, in which the two phenotypes coexist in the tumour, can be calculated if we assume that for an equilibrium to exist, the payoff of one strategy is equal to the payoff of the other strategy, thus $pE(m,m) + (1-p)E(m,p) = pE(p,m) + (1-p)E(p,p)$, where $p$ is the probability of a cell having the motile phenotype and \textit{E(x,y)} is the payoff for a cell with phenotype \textit{x} when meeting a cell with phenotype \textit{y}. The values of \textit{E} can be extracted from  table \ref{tab:payoffTable} and thus, the probability \textit{p} of adopting strategy motility in equilibrium varies with \textit{b} and \textit{c} (with \textit{c} not more than half the base payoff) as shown in the following equation:

\begin{equation}
p = \frac{b-2c}{b-c}
\label{eq:p}
\end{equation}

\begin{figure}
	\centering
		 \includegraphics[scale=0.4]{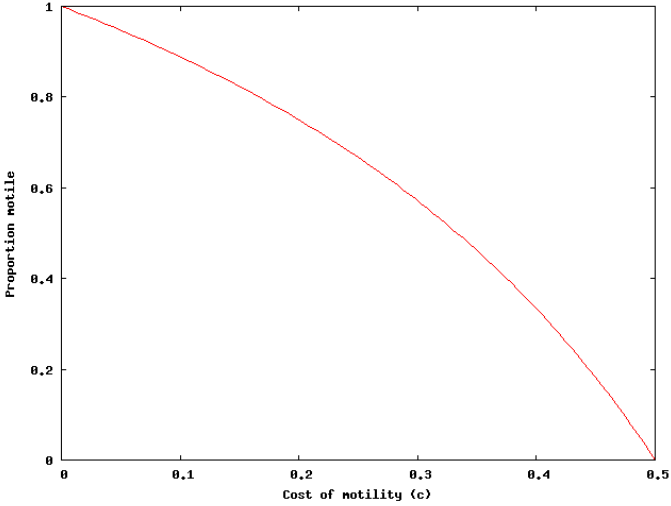}
	\caption{The graph shows  the results of analysing the game theoretical model. The X axis shows the value of the cost of motility and the Y axis shows the proportion of the population with the motile phenotype.}
	\label{fig:pm}
\end{figure}

Fig. \ref{fig:pm} shows the proportion of motile phenotypes in the population when the base payoff, $b$, is fixed. It can be proven that in games with only two pure strategies there is always one stable mixed strategy and the composition of pure strategies in it corresponds to a polymorphism \cite{Maynard:1982}.

\section{The cellular automaton model}
\begin{figure} 
\begin{minipage}{1\hsize}
	\subfigure[]{\includegraphics[width=0.5\hsize]{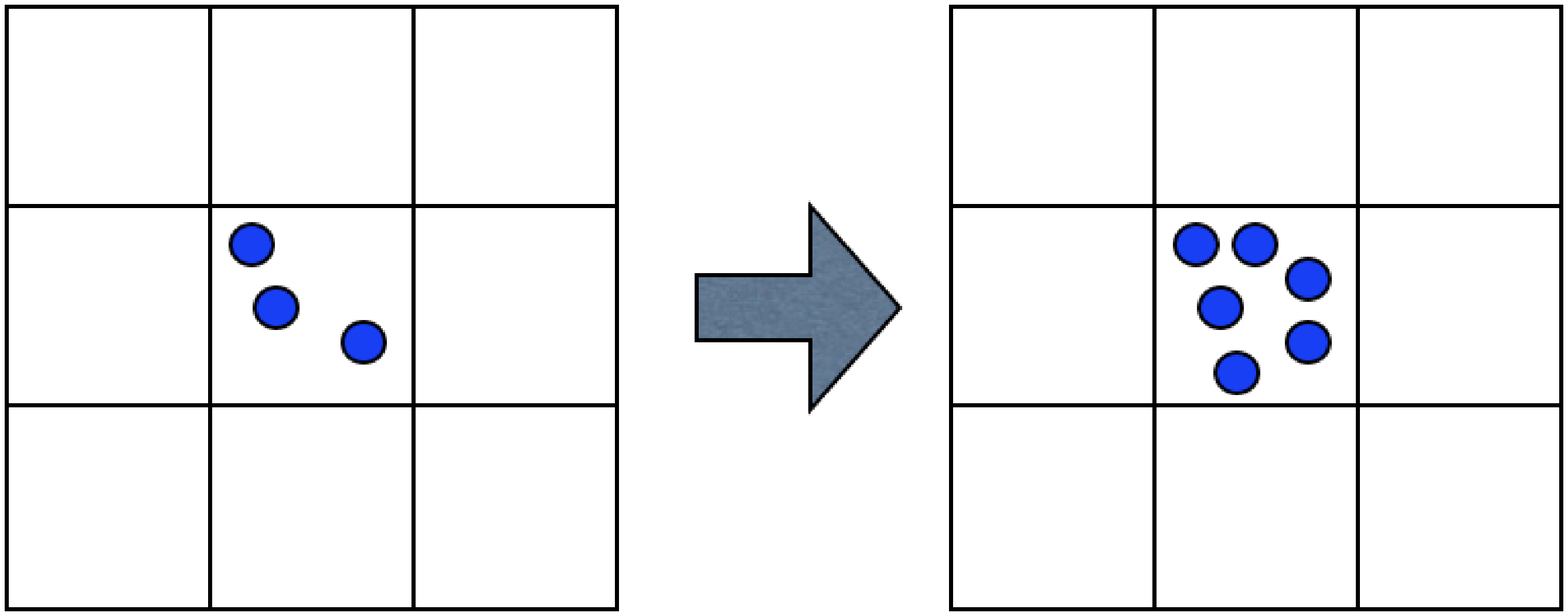}}\hfill
	\subfigure[]{\includegraphics[width=0.5\hsize]{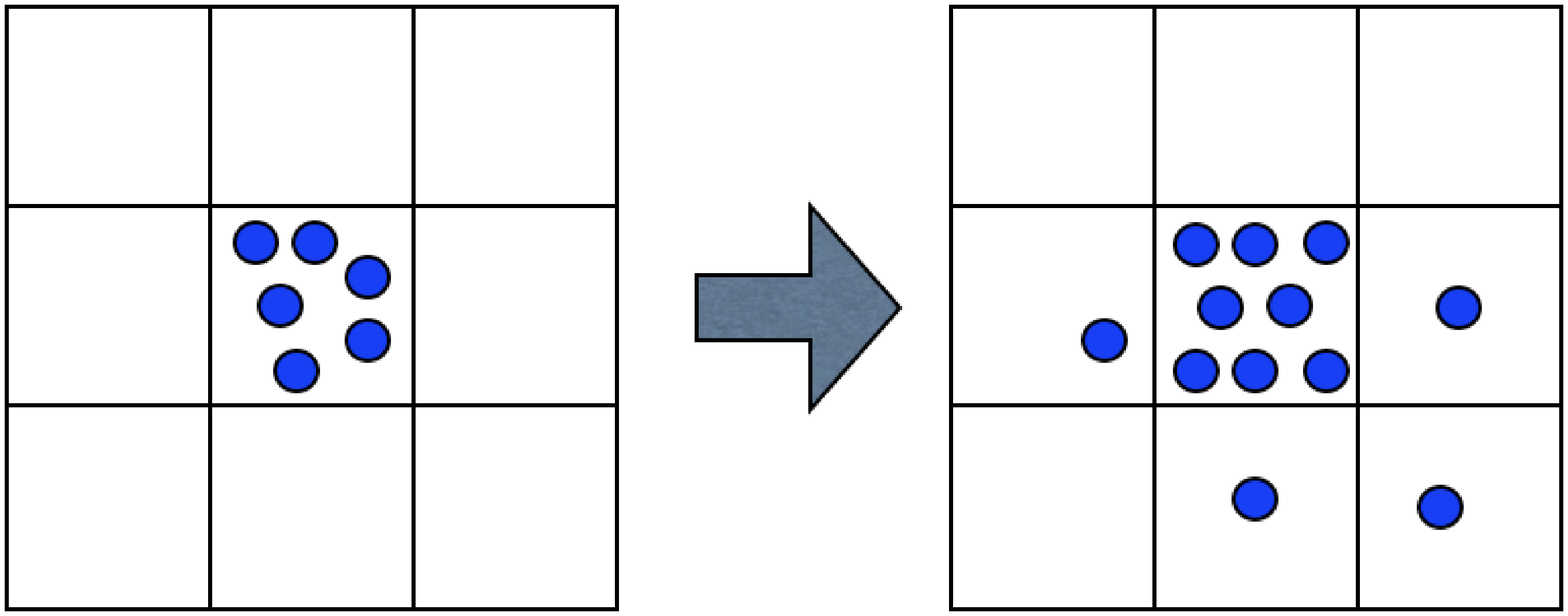}}
\end{minipage}
\caption{Behaviour of proliferative cells. (a) In the example each lattice site can host up to 8 tumour cells. When the lattice site contains enough space for the proliferative cells to produce their offspring in their current location they will do so. (b) If the lattice site can not contain the new offspring then it will be created in the neighbouring site with the most available space. In the example the first two cells selected have enough space in their site to create their offspring. The remaining ones have to select a neighbouring site and since empty ones are preferred then each cell choses a different neighbouring site. }
\label{fig:CAprolif}
\end{figure}

\begin{figure} 
\begin{minipage}{1\hsize}
	\subfigure[]{\includegraphics[width=0.5\hsize]{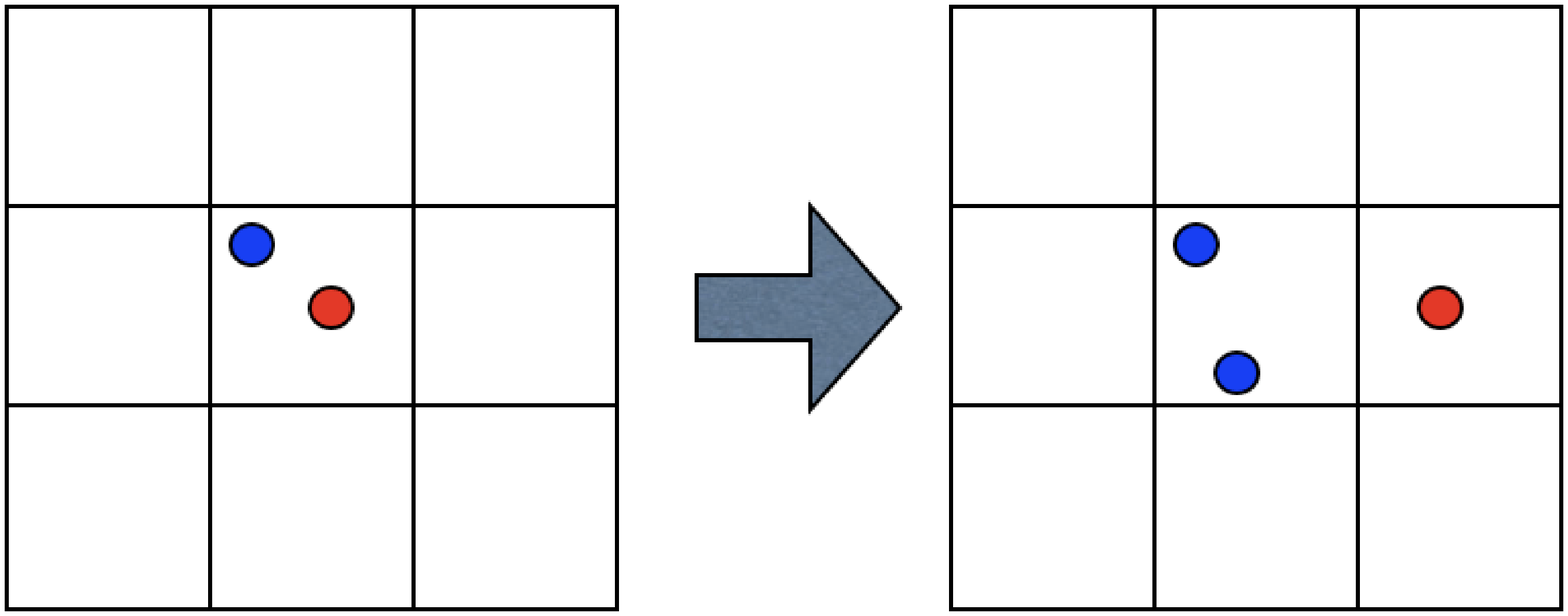}}\hfill
	\subfigure[]{\includegraphics[width=0.5\hsize]{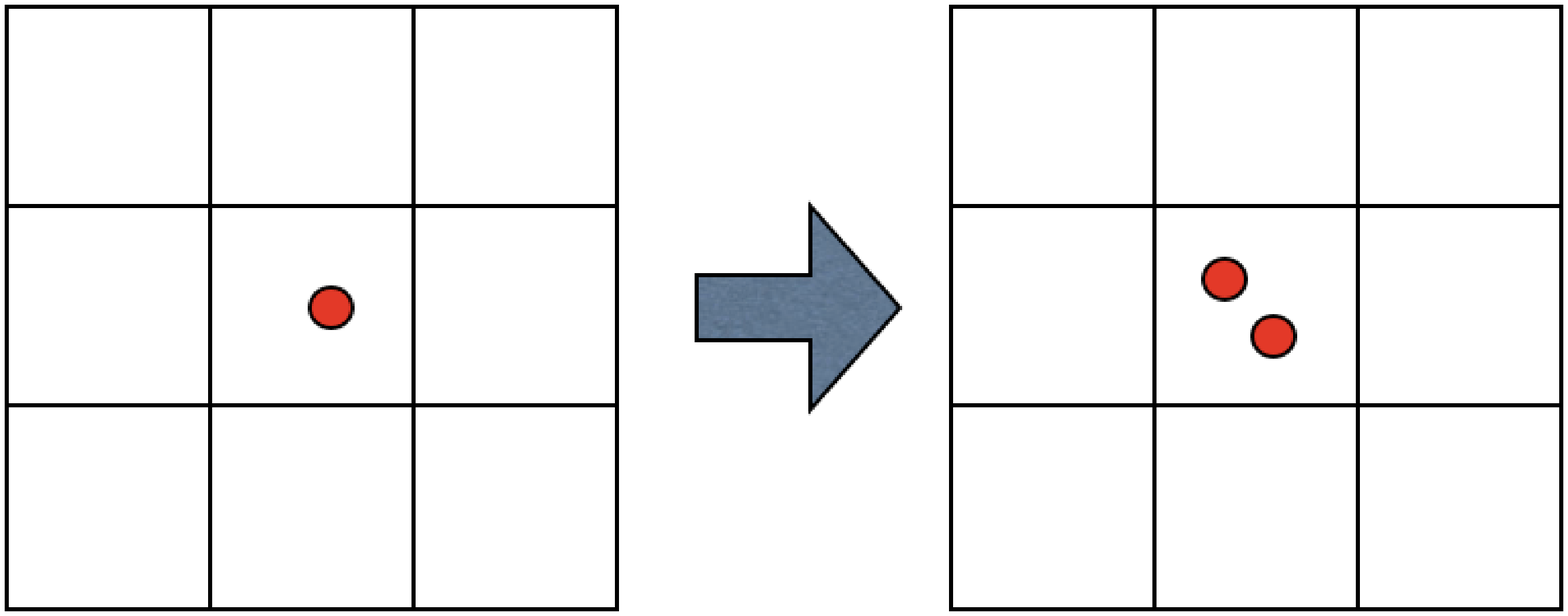}}
\end{minipage}
\caption{Behaviour of motile cells. (a) In this example motile cells are red and proliferative cells are blue. If a motile cell is not alone in the lattice site then it will move to a neighbouring site with the smallest cell density. This will leave all the site resources to the proliferative cell. (b) On the other hand if it does not have to share the site with any other tumour cell then it will divide. In the following time step, at least one of them will have to move to a different location if there is any site with a lower cell count in the neighbourhood. }
\label{fig:CAmot}
\end{figure}

The GT model does not take into account the influence of space in the dynamics of tumour evolution. For that reason we have introduced a cellular automaton (CA) with the aim of producing a spatial model in which the behaviour of proliferative and motile cells can be investigated. Space in the CA is characterised by a 1000 x 1000 lattice in which each site can host one or more tumour cells. The carrying capacity of a lattice site corresponds to the area in which one cell can affect another one. Each lattice site also contains nutrients which are needed for cell proliferation. Nutrients are assumed to be homogeneously distributed in the lattice and replenished to a given constant value each time step. Thus the resources that determine cell proliferation are available space and nutrients. Cells interact and share resources only with other cells in the same lattice site. This very simple microenvironment mimics the one that is implictly assumed in the GT model.
The behaviour of the tumour cell types was designed to resemble, as closely as possible, that of the two phenotypes of the GT model. It features proliferative and motile cells. The former do not move even if that means sharing an ever decreasing amount of resources. The latter divide only when they do not have to share the available resources with no other cell and try to move to a better location otherwise. The precise behaviour of the different phenotypes is the following. The cells with the proliferative phenotype divide if there are enough nutrients in the lattice site. The new offspring will be created in the same site as the progenitor unless the population of cells at that site has exceeded the maximum carrying capacity. In the last case the offspring is created in a neighbouring location having the lowest density of cells (see figure \ref{fig:CAprolif}).
Motile cells will never share a lattice site with another cell if there is more available space in a neighbouring location. Thus, motile cells always move to a neighbouring site if the density of the new location is lower than that of the current location. If more than one neighbouring location is equally suitable then the new location will be selected randomly among them. On the other hand if the location is suitable and there are enough nutrients then the cell will produce offspring in the same location as the parent (see figure \ref{fig:CAmot}). 

The update of the CA is as follows: 
\begin{itemize}
\item If cell is motile and there are neighbours:
	\begin{itemize}
		\item Look for the neighbouring lattice site with the smallest cell count.
		\item Move to new location if cell count is smaller than in current location.
	\end{itemize}
	\item Otherwise:
	\begin{itemize}
		\item If there is enough space in the current lattice site then divide.
		\item else
		\begin{itemize}
			\item Look for the lattice site with the smallest number of cells.
			\item If chosen site has enough space produce offspring.
			\item If offspring has been produced then, with a small probability, mutate new cell to motile.
		\end{itemize}
	\end{itemize}
\end{itemize}

Every time step each of the tumour cells is chosen to have its programme executed. The order in which the cells are called is randomly selected in each time step to avoid potential biases. The simulations start with a group of proliferative cells placed in the centre of the lattice (circumference with radius 20 lattice sites). Every time proliferative cells divide there is a small chance ($10^{-5}$) that there will be a mutation conferring the offspring the motile phenotype. Motile cells on the other hand will never produce proliferative cells when dividing. The reasoning for this is that the mutations in our model are not necessarily genetical but all have an effect on the phenotype. Whereas cells in tumour in an early stage of development tend to be proliferative, the acquisition of motility and invasiveness is a necessary capability that may emerge as a result of a number of genetic and epigenetic mutations \cite{hanahan:2000}. Hence, it is very unlikely that a motile cell could produce offspring that would accumulate the necessary mutations to disable this capability.

\section{Simulations and analysis}

\begin{figure} 
\begin{minipage}{1\hsize}
	\subfigure[ts=0]{\includegraphics[width=0.3\hsize]{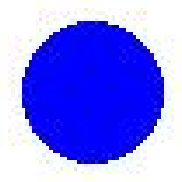}}\hfill
	\subfigure[ts=100]{\includegraphics[width=0.3\hsize]{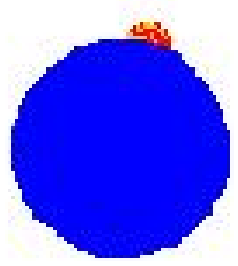}}\hfill
	\subfigure[ts=200]{\includegraphics[width=0.3\hsize]{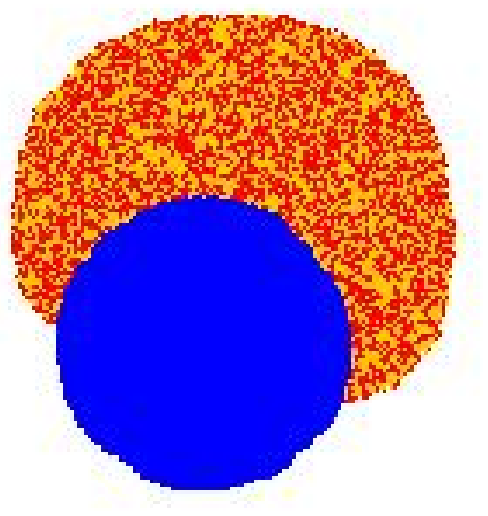}}\hfill
	\subfigure[ts=300]{\includegraphics[width=0.3\hsize]{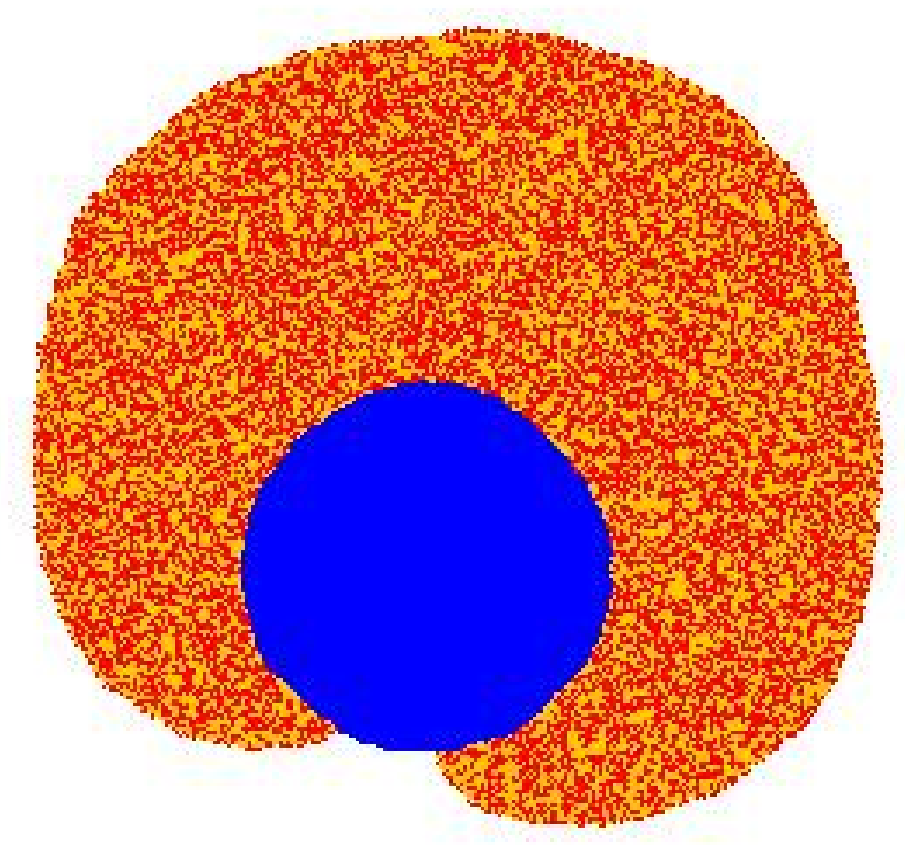}}\hfill
	\subfigure[ts=400]{\includegraphics[width=0.3\hsize]{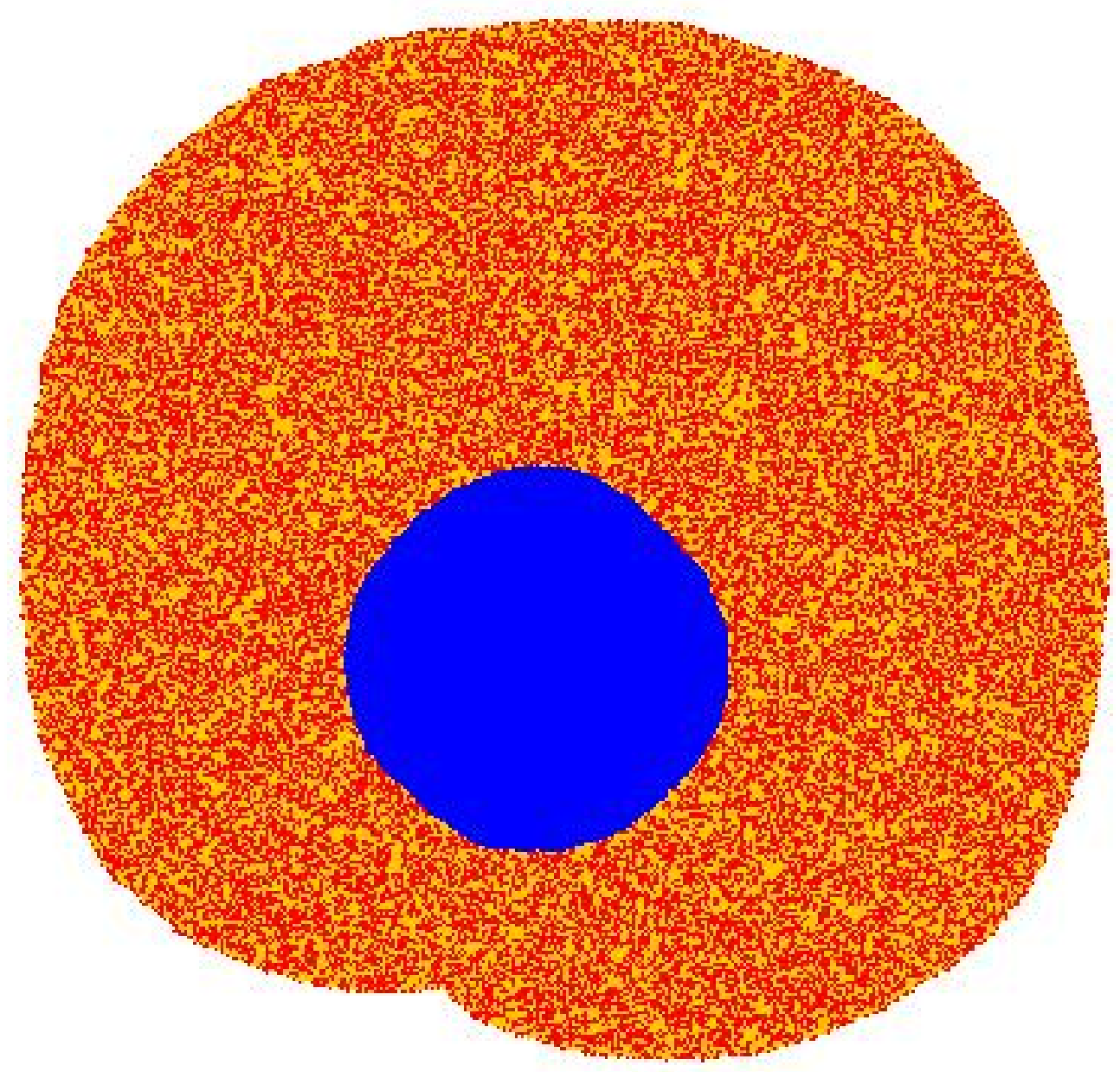}}\hfill 
	\subfigure[ts=500]{\includegraphics[width=0.3\hsize]{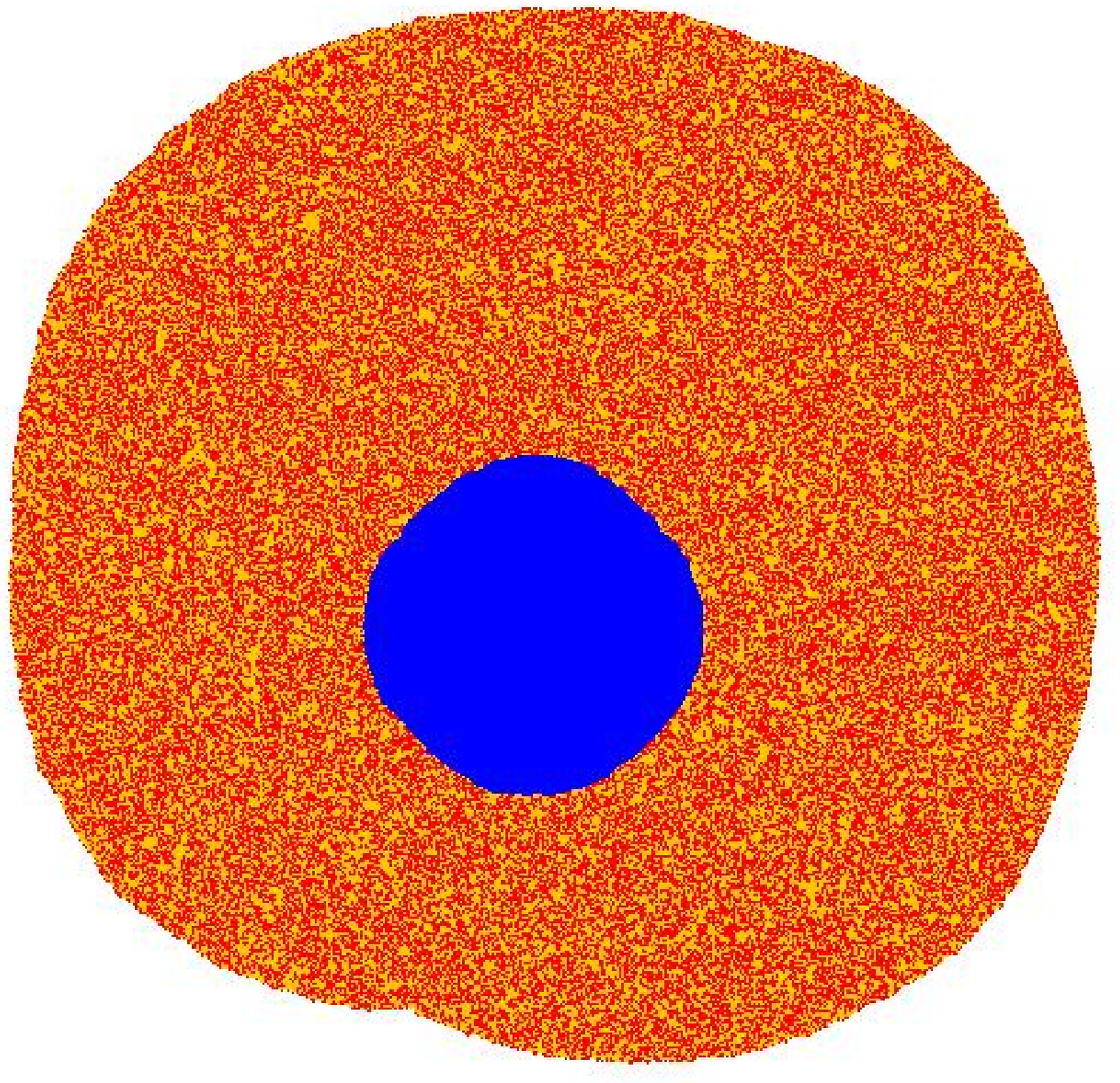}}
\end{minipage}
\caption{Result of a sample simulation produced by the CA model. In this particular simulation each lattice site can host up to 10 tumour cells. Every time step only one unit of nutrients is replenished in each site. These circumstances reduce the cost of motility and thus select for the motile phenotype. Lattice sites which contain exclusively proliferative cells are marked in blue, those that contain only motile cells are red and finally those that contain more than one type of cell are orange.}
\label{fig:CAsimul}
\end{figure}

The CA model was used to produce several simulations (e.g.: figure \ref{fig:CAsimul}), each one lasting 5000 timesteps, in which different values of nutrient replenishment and capacity of the lattice sites were tested. 

\begin{figure}
	\centering
		 \includegraphics[scale=1]{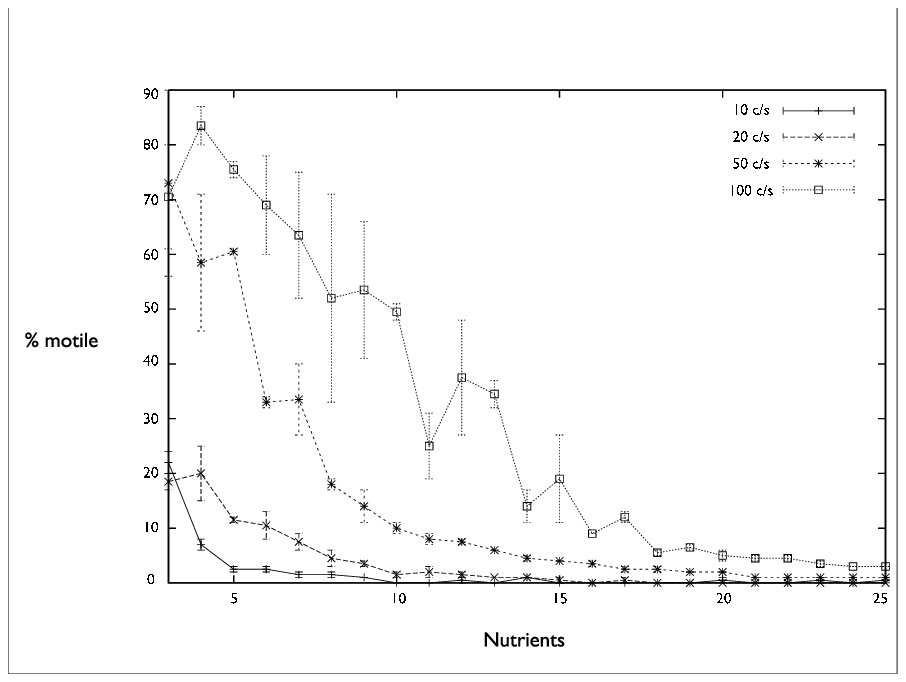}
	\caption{The graph shows the proportion of motile cells (Y axis) that result from using different amount of nutrients per time step (X axis) after running the CA model for 500 time steps. The results are for different  carrying capacities of each lattice site (10, 20, 50 and 100 cells/site). In all cases increasing the amount of nutrients supplied each time step reduces the proportion of motile tumour cells.}
	\label{fig:cares}
\end{figure}

Figure \ref{fig:cares} shows the results of the simulations. The curves represent the proportion of motile cells that results for values of nutrient replenishment between 1 and 25. Each line represents a different lattice site carrying capacity of 10, 20, 50 and 100 cells per site.

The results from both analysis and simulations support current medical knowledge: if the cost of motility is zero then the motility phenotype will reach fixation and spread through the tumour population. If the cost of motility increases, so does the proportion of proliferative cells in the population. The proportion of motile cells in the simulations depends on the carrying capacity and the amount of nutrients. Simulations in which lattice sites can host many cells but nutrient availability is scarce, the cost of motility in relation to the cost of remaining in the same place is quite low and thus yield a high proportion of motile cells. On the other hand when lattice sites can carry just a few cells forcing proliferative cells to produce their offspring in neighbouring sites then the relative cost of motility increases and likewise the proportion of proliferative cells.

In the CA model the ratio of nutrients per lattice site capacity gives a measure of the cost of motility ($c$) relative to the maximum fitness benefit ($b$). When the ratio of nutrients per lattice site capacity is high then the relative cost of motility will be high too. In these situations there are plenty of nutrients so that proliferative cells normally have enough resources to divide in every time step and thus out-compete motile cells. On the other hand if the ratio of nutrients per cell site is low then many proliferative cells will not have enough nutrients and might have less opportunities to produce offspring than cells with the motile phenotype.

\section{Conclusions}
We have introduced a game theoretical model that allows the study of the emergence of invasion and motility in tumours made of proliferative cells. We have also shown a CA model designed to study how spatial considerations could alter the results yielded by the GT analysis. The existence of polymorphic tumours has been found in other GT models \cite{tomlinson:1997b} and the results from both, our GT analysis and the CA-based simulations, confirm that there are many circumstances under which different strategies could be present in the tumour cell population at a given time.
Many simplifications were made in order to get a model tractable enough to be simulated and analysed. It is clear that more than two phenotypes are possible at this stage of carcinogenesis and that tumour cells usually interact in a more complex fashion than the one used in this study. The microenvironment was also kept as simple as possible. It is clear that more complex microenvironments (with differences in the concentration of e.g. pH, oxygen, ECM and nutrients) should have an influence of the fitness of the two phenotypes. Many of these simplifications are necessary and similar to those adopted by other researchers \cite{tomlinson:1997a,tomlinson:1997b,bach:2001,bach:2003,mansury:2006}. This fact does not detract merit from the potential of the model to test the ideas behind the word models usually put forward by oncologists \cite{hanahan:2000}. 
The simulations also support a therapy that would raise the cost of motility with respect to the base benefit to reduce the proportion of tumour cells capable of motility. Thus, by increasing the relative fitness advantage of non-motile phenotypes increases the probability of the tumour remaining benign. Increasing the cost of motility could be achieved in at least two different ways. One option would be to design therapies to hinder the detachment of tumour cells from the extra-cellular matrix by means of downregulation of integrins \cite{friedl:2003}. Alternatively, the model suggests the counter-intuitive idea that it should be possible to increase the fitness of proliferative cells over motile cells by increasing the amount of nutrients in the neighbourhood of the tumour. Increasing the amount of nutrients (equ. \ref{eq:p}) produces the same effect as acting directly on the cost of motility. It seems reasonable to think that increasing the amount of nutrients would act as a disincentive for a tumour cell to become motile. This is not necessarily a therapy that would work in all cases: there are benign tumours that are life threatening even if they do not become invasive. Despite that, there are many cases in which a growing but non aggressive tumour will have a much better prognosis than a smaller but invasive one. In those cases this therapy could help the patient by producing a tumour environment that selects for less aggressive phenotypes. 

\section*{Acknowledgements}
The authors would like to thank Michael K\"ucken from Technische Universit\"at Dresden for his helpful suggestions in the preparation of this manuscript. The work in this paper was supported by funds from the Marie Curie Network EU-RTD-IST-2001-38923. We also acknowledge the support provided by the systems biology network HepatoSys of the German Ministry for Education and Research through grant 0313082C. AD is  a member of the  DFG-Center for Regenerative Therapies Dresden - Cluster of Excellence - and gratefully acknowledges support by the Center.
\bibliography{cagt}

\begin{thebibliography}{10}

\bibitem{nowell:1976}
P.C. Nowell.
\newblock The clonal evolution of tumor cell populations.
\newblock {\em Science}, 4260, 194:23--28, 1976.

\bibitem{hanahan:2000}
D.~Hanahan and R.~Weinberg.
\newblock The hallmarks of cancer.
\newblock {\em Cell}, 100:57--70, Jan. 2000.

\bibitem{mareel:1998}
M.~Mareel and F.~Van~Roy.
\newblock The human {E}-cadherin/catenin complex: a potent invasion and tumor
  supressor, 1998.

\bibitem{foty:2004}
R.~A. Foty and M.~S. Steinberg.
\newblock Cadherin-mediated cell-cell adhesion and tissue segregation in
  relation to malignancy.
\newblock {\em Int. J. Dev. Biol.}, 48:397--409, 2004.

\bibitem{Merlo:2006}
L.~Merlo, J.~Pepper, B.~Reid, and C.~Maley.
\newblock Cancer as an evolutionary and ecological process.
\newblock {\em Nat. Rev. Cancer}, 6:924--935, December 2006.

\bibitem{gatenby:2003}
R.~Gatenby and P.~Maini.
\newblock Cancer summed up.
\newblock {\em Nature}, 421:321, January 2003.

\bibitem{Neumann:1953}
J.~von Neumann and O.~Morgernstern.
\newblock {\em Theory of games and economic behaviour}.
\newblock Princeton University Press, Princeton, NJ, 1953.

\bibitem{Merston-Gibbons:2000}
Michael Merston-Gibbons.
\newblock {\em An introduction to game-theoretic modelling}.
\newblock American Mathematical Society, 2 edition, 2000.

\bibitem{Maynard:1982}
J.~Maynard~Smith.
\newblock {\em Evolution and the theory of games}.
\newblock Cambridge University Press, Cambridge, 1982.

\bibitem{Hofbauer:1998}
J.~Hofbauer and K.~Sigmund.
\newblock {\em Evolutionary games and population dynamics}.
\newblock Cambridge University Press, Cambridge, 1998.

\bibitem{Basanta:2008}
D.~Basanta and A.~Deutsch.
\newblock A game theoretical perspective on the somatic evolution of cancer.
\newblock In Selected topics on cancer modelling: genesis, evolution, inmune
  competition, therapy (N. Bellomo, M. Chaplain and E. De Angelis Eds.).
  Birkhauser, Boston, 2008.

\bibitem{tomlinson:1997a}
I.P.M. Tomlinson.
\newblock Game theory models of interactions between tumour cells.
\newblock Eur. Jour. Cancer, Vol 33, N9, pp. 1495-1500, 1997.

\bibitem{tomlinson:1997b}
I.P.M. Tomlinson and W.~F. Bodmer.
\newblock Modelling the consequences of interactions between tumour cells.
\newblock {\em Brit. Jour. Cancer}, 75(2):157--60, 1997.

\bibitem{bach:2001}
L.A. Bach, S.~M. Bentzen, J.~Alsner, and F.~B. Christiansen.
\newblock An evolutionary game model of tumour cell interactions: possible
  relevance to gene therapy.
\newblock {\em Eur. J. Cance.}, 37:2116--2120, 2001.

\bibitem{bach:2003}
L.A. Bach, D.~J.~T.. Sumpter, J.~Alsner, and V.~Loeschke.
\newblock Spatial evolutionary games of interaction among generic cancer cells.
\newblock {\em J. Theor. Med.}, 5(1):47--58, 2003.

\bibitem{gatenby:2003a}
R.~Gatenby and T.~Vincent.
\newblock An evolutionary model of carcinogenesis.
\newblock {\em Cancer Res.}, 63:6212--6220, October 2003.

\bibitem{Gatenby:2005}
R.~Gatenby, T.~Vincent, and R.~Gillies.
\newblock Evolutionary dynamics in carcinogenesis.
\newblock {\em MATH MOD METH APPL S}, 15(11):1619--1638, 2005.

\bibitem{gatenby:2003b}
R.~Gatenby and T.~Vincent.
\newblock Application of quantitative models from population biology and
  evolutionary game theory to tumor therapeutic strategies.
\newblock {\em Mol. Cancer Ther.}, 2:919--927, 2007.

\bibitem{mansury:2006}
Y.~Mansury, M.~Diggory, and T.~S. Deisboeck.
\newblock Evolutionary game theory in an agent based brain tumor model:
  exploring the genoype phenotype link.
\newblock {\em J. Theo. Biol.}, 238:146--156, 2006.

\bibitem{Axelrod:1981}
R.~Axelrod and W.~Hamilton.
\newblock The evolution of cooperation.
\newblock {\em Science}, 211:1390--1396, 1981.

\bibitem{Axelrod:2006}
R.~Axelrod, D.~Axelrod, and K.~Pienta.
\newblock Evolution of cooperation among tumor cells.
\newblock {\em PNAS}, 103(36):13474--79, Sept. 2006.

\bibitem{Neumann:1966}
J.~von Neumann.
\newblock {\em Theory of self-reproducing automata}.
\newblock University of Illinois Press, 1966.

\bibitem{deutsch:2005}
A.~Deutsch and S.~Dormann.
\newblock {\em ellular Automaton Modeling of Biological Pattern Formation:
  Characterization, Applications, and Analysis}.
\newblock Birkh{\"a}user, Boston, 2005.

\bibitem{Andreas-Deutsch-Editor:2007}
A.~Deutsch, L.~Brusch, H.~Byrne, G.~de~Vries, and H.~Herzel (Eds).
\newblock {\em Mathematical Modeling of Biological Systems, Volume I: Cellular
  Biophysics, Regulatory Networks, Development, Biomedicine, and Data Analysis
  (Modeling ... in Science, Engineering and Technology)}.
\newblock Birkhauser Boston, 2008.

\bibitem{Moreira:2002}
J.~Moreira and A.~Deutsch.
\newblock Cellular automaton models of tumor development: a critical review.
\newblock {\em Adv. Compl. Sys.}, 5(2/3):247, 2002.

\bibitem{Kansal:2000}
A.~R. Kansal, S.~Torquato, E.~A. Chiocca, and T.~S. Deisboeck.
\newblock Emergence of a subpopulation in a computational model of tumor
  growth.
\newblock {\em J. Theor. Biol.}, 207(431-441), 2000.

\bibitem{Ribba:2004}
B.~Ribba, T.~Alarcon, K.~Marron, P.~Maini, and Z.~Agur.
\newblock The use of hybrid cellular automaton models for improving cancer
  therapy.
\newblock In Proceedings, Cellular Automata: 6th International Conference on
  Cellular Automata for Research and Industry, ACRI 2004, 2004.

\bibitem{Wurzel:2005}
M.~Wurzel, K.~L. Schaller, M.~Simon, and A.~Deutsch.
\newblock Brain cancer invasion of brain tissue: guided by a prepattern.
\newblock {\em J. Theor. Medic.}, 6(1):21--31, 2005.

\bibitem{Hatzikirou:2005}
H.~Hatzikirou, A.~Deutsch, C.~Schaller, M.~Simon, and K.~Swanson.
\newblock Mathematical modelling of glioblastoma development.
\newblock {\em Math. Mod. Meth. Appl. Sci.}, pages 1779--1794, 2005.

\bibitem{Hatzikirou:2007b}
H.~Hatzikirou and A~Deutsch.
\newblock Cellular automata as microscopic models of cell migration in
  heterogeneous environments.
\newblock {\em Curr. Top. Dev. Biol.}, 81, 2007.

\bibitem{Alarcon:2003}
T.~Alarcon, H.~M. Byrne, and P.~K. Maini.
\newblock A multiple scale model for tumour growth.
\newblock {\em J. Theor. Biol.}, 225:257--274, 2003.

\bibitem{Merks:2006}
R.~M.~H. Merks and J.~A. Glazier.
\newblock Dynamic mechanisms of blood vessel growth.
\newblock {\em Nonlinearity}, 19:1--10, 2006.

\bibitem{Spencer:2006}
S.~Spencer, R.~Gerety, K.~Pienta, and S.~Forrest.
\newblock Modeling somatic evolution in tumorigenesis.
\newblock {\em PLOS Computational Biology}, 2(8):939--947, August 2006.

\bibitem{Anderson:2006}
A.~Anderson, A.~Weaver, P.~Cummings, and V.~Quaranta.
\newblock Tumor morphology and phenotypic evolution driven by selective
  pressure from the microenvironment.
\newblock {\em Cell}, 127:905--915, December 2006.

\bibitem{giese:1996}
A.~Giese, M.~Loo, N.. Tran, S.~W. Haskett, and M.~E. Berens.
\newblock Dichotomy of astrocytoma migration and proliferation.
\newblock {\em Int. J. Cance.}, 67:275--282, 1996.

\bibitem{giese:2003}
A.~Giese, R.~Bjerkvig, M.~E. Berens, and M.~Westphal.
\newblock Cost of migration: invasion of malignant gliomas and implications for
  treatment.
\newblock {\em J. Clin. Oncol.}, 21:1624--1636, 2003.

\bibitem{friedl:2003}
P.~Friedl and K.~Wolf.
\newblock Tumour-cell invasion and migration: diversity and escape mechanisms.
\newblock {\em Nat Rev Cancer.}, 3(5):362--74, 2003.

\end{thebibliography}
\bibliographystyle{unsrt}
\printindex
\end{document}